\documentclass{Interspeech}

\usepackage{graphicx}
\usepackage{subcaption}
\usepackage{booktabs}
\usepackage{siunitx}
\usepackage{amssymb}
\usepackage{rotating}
\usepackage{makecell}
\usepackage{hyperref}
\usepackage{multirow}

\interspeechcameraready

\title{FlowSE: Efficient and High-Quality Speech Enhancement via Flow Matching}

\author[affiliation={1}\dagger]{Ziqian}{Wang}
\author[affiliation={1}\dagger]{Zikai}{Liu}
\author[affiliation={1}]{Xinfa}{Zhu}
\author[affiliation={1}]{Yike}{Zhu}
\author[affiliation={1}]{Mingshuai}{Liu}
\author[affiliation={2}]{Jun}{Chen}
\author[affiliation={2}]{Longshuai}{Xiao}
\author[affiliation={2}]{Chao}{Weng}
\author[affiliation={1}\ast]{Lei}{Xie}

\affiliation{Audio, Speech and Language Processing Group (ASLP@NPU), School of Software}{Northwestern Polytechnical University, Xi'an}{China}
\affiliation{}{Huawei Technologies}{China}
\email{zq\_wang@mail.nwpu.edu.cn, liuzikai@mail.nwpu.edu.cn}

\usepackage{comment}

\begin{document}

\maketitle

\begin{abstract}
Generative models have excelled in audio tasks using approaches such as language models, diffusion, and flow matching. However, existing generative approaches for speech enhancement (SE) face notable challenges: language model-based methods suffer from quantization loss, leading to compromised speaker similarity and intelligibility, while diffusion models require complex training and high inference latency. To address these challenges, we propose \textbf{FlowSE}, a \textbf{flow}-matching-based model for \textbf{SE}. Flow matching learns a continuous transformation between noisy and clean speech distributions in a single pass, significantly reducing inference latency while maintaining high-quality reconstruction. Specifically, FlowSE trains on noisy mel spectrograms and optional character sequences, optimizing a condition flow matching loss with ground-truth mel spectrograms as supervision. It implicitly learns speech’s temporal-spectral structure and text-speech alignment. During inference, FlowSE can operate with or without textual information, achieving impressive results in both scenarios, with further improvements when transcripts are available. Extensive experiments demonstrate that FlowSE significantly outperforms state-of-the-art generative methods, establishing a new paradigm for generative-based SE and demonstrating the potential of flow matching to advance the field. Our code, pre-trained checkpoints, and audio samples are available at \url{https://github.com/Honee-W/FlowSE/}.

\end{abstract}

\renewcommand{\thefootnote}{\fnsymbol{footnote}}  % 用符号编号
\footnotetext[2]{Equal contribution.\hspace{1em}\footnotemark[1]Corresponding author.}

\section{Introduction}

Speech enhancement (SE) aims to recover clean speech from noisy signals, playing a vital role in applications such as telecommunications, hearing aids, and speech recognition front ends. While traditional deterministic methods can attenuate noise, they often struggle to preserve speech naturalness under challenging conditions. Recent advances in generative modeling offer powerful alternatives for high-fidelity SE.

Generative SE methods have primarily pursued two directions. One direction leverages language model (LM) frameworks~\cite{wang2024selm, li2024masksr}, which use discrete representations obtained via quantization to generate clean speech. While these methods can be effective, the quantization process often leads to information loss~\cite{van2017neural, zeghidour2021soundstream, defossez2022encodec}, resulting in artifacts that compromise speaker similarity and intelligibility. The other direction employs diffusion models~\cite{lu2022conditional, welker2022speech, lemercier2023storm}, which iteratively refine noisy inputs through a stochastic denoising process. Despite impressive performance, diffusion-based approaches are computationally intensive and exhibit high latency, limiting their real-time applicability.

Concurrently, \emph{flow matching}~\cite{lipman2023flow} has emerged as an efficient one-shot generative paradigm, learning a continuous velocity field that transports simple noise distributions to complex data distributions. Flow matching has powered state-of-the-art results in speech synthesis~\cite{mehta2024matcha, eskimez2024e2, chen2024f5}, speech enhancement~\cite{nugraha2020flow, jung2024flowavse}, and sound separation~\cite{yuan2025flowsep}, demonstrating its ability to combine high fidelity with fast sampling.

Motivated by these advances, we propose FlowSE, among the first SE frameworks to employ rectified flow matching with a latent Diffusion Transformer (DiT) backbone. Unlike prior models that relied on U-Net or VAE architectures, FlowSE’s DiT network better captures long-range dependencies across time and frequency, enhancing the reconstruction of complex speech patterns.FlowSE learns a single-pass mapping from noisy to clean mel spectrograms, avoiding LM quantization and diffusion’s iterative complexity. FlowSE trains on noisy mel spectrograms and optional transcript sequences, optimizing a condition flow matching loss with ground-truth mel spectrograms as supervision. It implicitly learns speech’s temporal-spectral structure and text-speech alignment. During inference, FlowSE can operate with or without textual information, achieving impressive results in both scenarios, with further improvements when transcripts are available.

In summary, we introduce FlowSE, a rectified flow matching framework with a DiT backbone for efficient, high-fidelity SE that preserves speaker identity. Extensive experiments demonstrate that FlowSE significantly improves speech quality, intelligibility, and speaker similarity while reducing inference latency compared to existing generative SE methods.

\section{Related Work}
\label{sec:related_work}

\subsection{Generative-Based Speech Enhancement}
Generative approaches have recently become a prominent focus in speech enhancement (SE), effectively restoring clean speech signals from noisy inputs. Traditional deep learning-based SE methods, such as convolutional neural networks (CNNs) and recurrent neural networks (RNNs), primarily focus on deterministic speech reconstruction~\cite{luo2019conv, defossez2020real, hu2020dccrn}. In contrast, generative models learn the underlying distribution of clean speech, enabling superior generalization to unseen noise conditions.

Early explorations of generative SE primarily centered on Variational Autoencoders (VAEs) and Generative Adversarial Networks (GANs)~\cite{ liu2020cp, kim2021multi}. VAEs demonstrate the potential of latent modeling but often struggle to produce sharp and high-fidelity outputs, leading to over-smoothed speech. GANs, on the other hand, improve perceptual quality by leveraging adversarial training but suffer from training instability and mode collapse, making them challenging to optimize for SE tasks.

Building on these early efforts, more recent work has explored LM approaches~\cite{wang2024selm, li2024masksr}. These methods encode speech into discrete tokens using a codec model and then apply an LM to predict clean speech from noisy inputs. While this approach effectively leverages powerful language models for speech restoration, it introduces a fundamental limitation: the quantization process results in information loss, leading to artifacts that degrade speaker similarity and intelligibility.

Another prominent generative SE framework is based on diffusion models~\cite{lemercier2023storm, richter2023speech}. These models iteratively refine speech signals through a stochastic denoising process, progressively reconstructing clean speech from noisy inputs. While diffusion-based SE methods achieve state-of-the-art performance under extreme noise conditions, their high computational cost and slow inference speed pose challenges for real-time deployment.

% \subsection{Flow Matching in Speech Generation}
% Flow matching~\cite{lipman2023flow} is a generative modeling paradigm that learns continuous-time probability flow trajectories to transform noise into structured data. Unlike diffusion models, which use a stochastic process, flow matching directly learns a time-dependent velocity field that smoothly maps data between noisy and clean distributions.

% Several works have successfully applied flow matching to speech tasks, particularly in text-to-speech (TTS). MATCHA-TTS~\cite{mehta2024matcha} uses a flow-matching model~\cite{lipman2023flow} conditioned by an input text. E2-TTS~\cite{eskimez2024e2} introduces a flow-matching-based approach for end-to-end TTS, demonstrating simple and efficient speech synthesis. F5-TTS~\cite{chen2024f5} further optimizes the process by incorporating a text encoder to refine the text representations and inference-time sway sampling strategy, improving performance and efficiency. 

\subsection{Flow Matching for Generative Modeling}
Flow matching, first introduced by Meta~\cite{lipman2023flow}, is a versatile generative modeling framework that learns continuous-time probability flows to transform a simple noise distribution into a complex target distribution. Unlike diffusion models—which rely on iterative stochastic denoising—flow matching directly learns a deterministic, time-dependent velocity field, enabling a smooth and efficient one-shot transformation from noise to data.

This paradigm has been successfully applied in speech generation tasks. In text-to-speech (TTS). MATCHA-TTS~\cite{mehta2024matcha} uses a flow-matching model conditioned by an input text. E2-TTS~\cite{eskimez2024e2} introduces a flow-matching-based approach for end-to-end TTS, demonstrating simple and efficient speech synthesis. F5-TTS~\cite{chen2024f5} further optimizes the process by incorporating a text encoder to refine the text representations and inference-time sway sampling strategy, improving performance and efficiency. Flow matching has also been explored for speech enhancement and separation. Nugraha et al. proposed GF-VAE, combining Glow flows with VAEs for spectrogram modeling and semi-supervised enhancement~\cite{nugraha2020flow}. Jung et al. introduced FlowAVSE, which conditions flow matching on both audio and visual cues via a U-Net architecture to perform single-step audiovisual enhancement~\cite{jung2024flowavse}. More recently, Yuan et al. presented FlowSep, using rectified flow matching in a VAE latent space for language-conditioned audio source separation~\cite{yuan2025flowsep}.

While these models demonstrate the promise of flow-based approaches, they typically rely on U-Net or VAE backbones and, in some cases, visual inputs or latent space mappings. An audio-only, transformer-based framework that captures long-range dependencies and supports optional text conditioning remains unexplored.

\section{Proposed Approach}
\label{sec:method}

% \subsection{Overall Framework}
% Speech enhancement aims to recover clean speech $x \in \mathbb{R}^T$ from its noisy observation $y \in \mathbb{R}^T$. Generative SE models such as diffusion-based methods~\cite{lemercier2023storm, richter2023speech} rely on iterative denoising processes, which lead to high computational costs and slow inference speed. To address these challenges, we introduce \textbf{FlowSE}, a speech enhancement model based on flow matching~\cite{lipman2023flow}, which enables efficient and high-quality waveform reconstruction.

% As illustrated in Figure~\ref{fig:flowse_overview}, FlowSE consists of three key components: a text encoder $\mathcal{T}$ that refines the text representation,
% making it easy to align with the speech, a flow-matching-based generative model $\mathcal{F}$ that estimates the probability flow between noisy and clean speech distributions.
% a vocoder $\mathcal{V}$ that reconstructs the enhanced speech waveform from the learned representations.

% Unlike diffusion models that require hundreds of denoising steps, flow-matching models directly estimate a continuous transformation, significantly reducing inference latency while maintaining high speech quality. The following sections provide a detailed formulation of the flow-matching process and the specific network architecture of FlowSE.

\subsection{Overall Framework}
Speech enhancement (SE) aims to recover clean speech $x \in \mathbb{R}^T$ from its noisy observation $y \in \mathbb{R}^T$, a task that requires effective modeling of both acoustic signals and any available auxiliary information. To address this challenge, we propose FlowSE, a novel approach that integrates a flow-matching framework to achieve efficient and high-quality speech reconstruction.

\begin{figure}[t]
    \centering
    \includegraphics[width=0.9\linewidth]{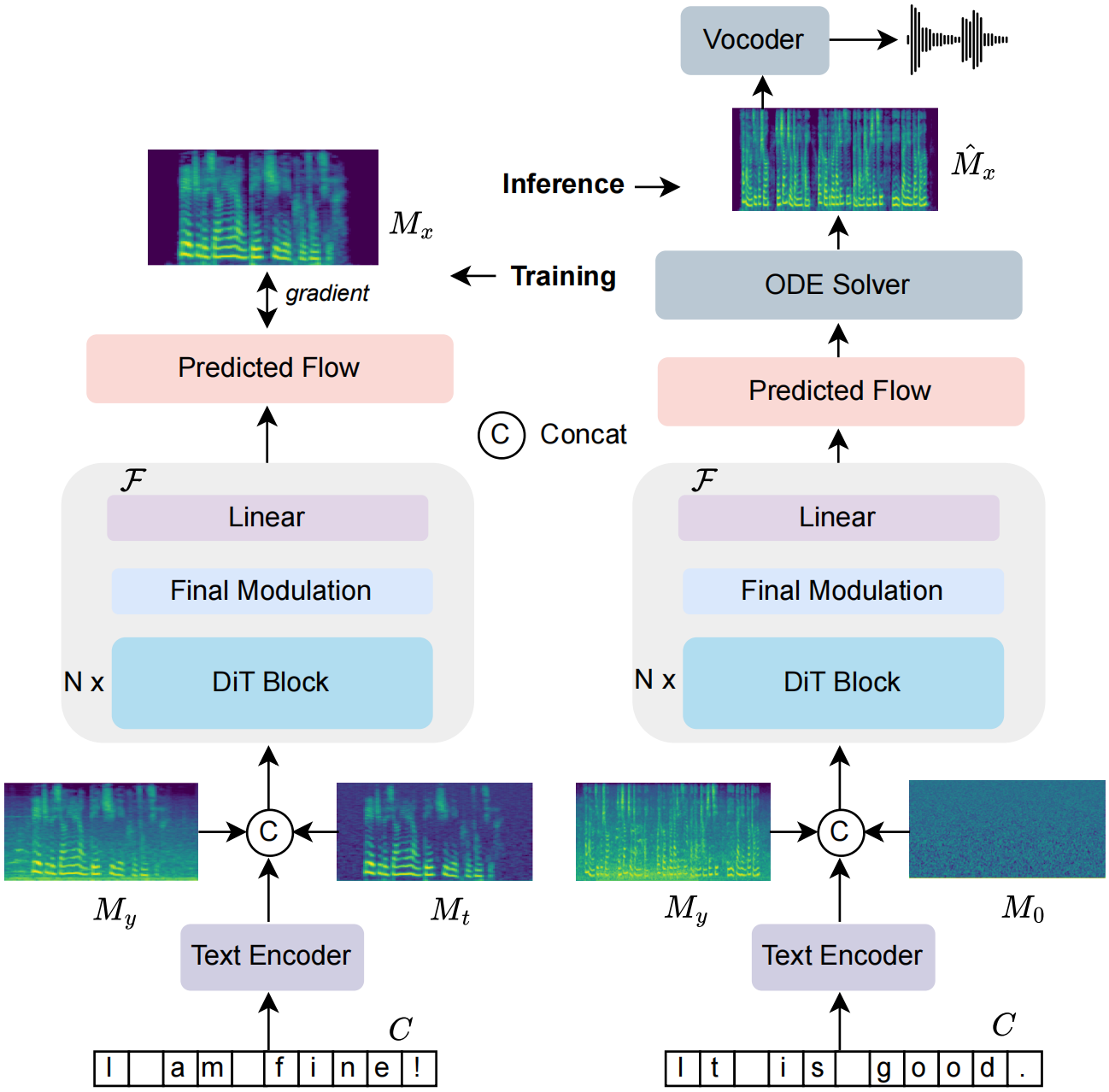}
    \caption{Overview of \textbf{FlowSE}. 
    During \textbf{training} (left), the model takes a noisy mel-spectrogram $M_y$, an interpolated mel-spectrogram $M_t$ (a mixture of clean speech and Gaussian noise), and an optional transcript $C$. The text is processed by a text encoder $\mathcal{T}$ and concatenated with the audio embeddings. The DiT-based flow model $\mathcal{F}$ predicts the probability flow, which is trained via gradient updates to reconstruct the clean speech mel-spectrogram $M_x$. 
    During \textbf{inference} (right), the model takes as input a noisy mel-spectrogram $M_y$, Gaussian noise $M_0$, and optional transcript $C$. The predicted flow guides the ODE solver to generate the enhanced mel-spectrogram $\hat{M}_x$, which is then converted to waveforms using a vocoder.}
    \label{fig:flowse_overview}
\end{figure}

As illustrated in Figure~\ref{fig:flowse_overview}, FlowSE comprises three key components. First, the text encoder $\mathcal{T}$ processes textual inputs to generate representations that align with the speech signal. Next, the flow-matching generative module $\mathcal{F}$ learns a continuous transformation mapping the noisy speech distribution to the clean one. Finally, the vocoder $\mathcal{V}$ reconstructs the enhanced waveform from these refined representations.

\vspace{-5pt}
\subsection{Flow Matching for Speech Enhancement}
\label{sec:flow_matching}

Flow matching~\cite{lipman2023flow} is a generative framework that uses continuous normalizing flows to map a source distribution $p(y)$ (e.g., noisy speech) to a target distribution $p(x)$ (e.g., clean speech). Unlike diffusion models, which require hundreds of iterative stochastic denoising steps, flow matching learns a deterministic ordinary differential equation(ODE) that considerably reduces the number of required iterations. Although solving this ODE may still require several evaluations, it typically involves far fewer steps (e.g., 10--20) than the hundreds of iterations used in diffusion models, thereby enabling near real-time inference. 

Specifically, it parameterizes a velocity field $v_\theta(z_t, t)$ to define a continuous transformation:
\begin{equation}
    \frac{d z_t}{d t} = v_\theta(z_t, t), \quad z_0 = y, \quad z_1 = x,
\end{equation}
where $z_0$ and $z_1$ denote the initial noisy speech and the final clean speech, respectively. During inference, the learned velocity field is integrated using an ODE solver:
\begin{equation}
    z_{t+\Delta t} = z_t + v_\theta(z_t, t)\,\Delta t.
\end{equation}
This approach reduces the need for stochastic sampling, enabling faster inference suitable for real-time SE.

\subsection{Network Architecture}
\label{sec:architecture}

FlowSE follows a three-stage processing pipeline: a mel-spectrogram encoder $\mathcal{E}$, a Flow Matching model $\mathcal{F}$ based on latent Diffusion Transformers (DiT)~\cite{peebles2023scalable}, and a vocoder $\mathcal{V}$ for waveform reconstruction.

\subsubsection{Mel-Spectrogram Encoder}
Unlike neural network-based encoders, FlowSE directly takes the mel-spectrogram representation of the input noisy speech. Given a raw waveform $y \in \mathbb{R}^{T}$, we first extract a mel-spectrogram $M_y \in \mathbb{R}^{F \times T'}$ using a standard short-time Fourier transform (STFT) and mel-filterbank transformation:
\begin{equation}
    % M_y = \text{Mel-STFT}(y),
    M_y = \mathcal{E}(y),
\end{equation}
where $F$ denotes the number of mel-frequency bins, and $T'$ is the number of frames. This representation serves as the input to the Flow Matching module.

% \subsubsection{Flow Matching with Latent Diffusion Transformers}
% To effectively model the transformation from noisy to clean speech, we adopt a flow-matching framework based on latent Diffusion Transformers (DiT)~\cite{peebles2023scalable}. The DiT operates directly on the mel-spectrogram domain, learning to parameterize the time-dependent velocity field $v_\theta(M_t, t)$, which guides the denoising process.

% Unlike conventional speech enhancement models that rely solely on acoustic features, FlowSE incorporates textual supervision by conditioning the model on corresponding transcripts. During training, the model is exposed to both noisy mel-spectrograms and their associated character sequences. To encourage implicit alignment learning, we randomly drop text inputs with a certain probability, forcing the model to develop robustness in both text-conditioned and text-free scenarios.

% Formally, given the intermediate mel-spectrogram state $M_t$ at time $t$, the DiT-based velocity model learns the mapping:
% \begin{equation}
%     v_\theta(M_t, t, C) = \mathcal{F}(M_t, t, \mathcal{T}(C)),
% \end{equation}
% where $C$ represents the optional character sequence derived from the transcript. When $C$ is provided, FlowSE leverages the textual context to enhance speech quality, particularly under low signal-to-noise ratio (SNR) conditions.

\subsubsection{Flow Matching with Latent Diffusion Transformers}
To effectively model the transformation from noisy to clean speech, we adopt a flow-matching framework based on latent Diffusion Transformers (DiT)~\cite{peebles2023scalable}. The DiT operates directly on the mel-spectrogram domain, learning to parameterize the time-dependent velocity field $v_\theta(M_t, t)$, which guides the transformation process.

During training, the model receives three types of inputs: (1) a noisy speech mel-spectrogram $M_y$, (2) an interpolated mel-spectrogram $M_t$ obtained by linearly combining the clean target $M_x$ with Gaussian noise, and (3) an optional text condition $C$ derived from the transcript. The velocity model is trained to estimate the probability flow that transports $M_t$ towards $M_x$ as time $t$ progresses.

At inference, the model takes as input (1) the mel-spectrogram of the noisy speech $M_y$, (2) a Gaussian noise initialization $M_0$, and (3) an optional text condition. The learned velocity field $v_\theta$ enables direct sampling using an ODE solver:

\begin{equation}
    M_{t+\Delta t} = M_t + v_\theta(M_t, t, C) \Delta t.
\end{equation}

To enhance generalization, textual supervision is incorporated by conditioning the model on transcripts. To encourage implicit alignment learning, text inputs are randomly dropped with a certain probability during training, ensuring robustness in both text-conditioned and text-free scenarios. When $C$ is provided, FlowSE employs a text encoder $\mathcal{T}$ to leverage textual context to improve speech quality, particularly under low signal-to-noise ratio (SNR) conditions.

Formally, given the intermediate mel-spectrogram state $M_t$ at time $t$, the DiT-based velocity model learns the mapping:

\begin{equation}
    v_\theta(M_t, t, C) = \mathcal{F}(M_t, t, \mathcal{T}(C)).
\end{equation}

\subsubsection{Vocoder for Waveform Reconstruction}
The final stage of FlowSE utilizes a pre-trained neural vocoder to reconstruct the waveform from the enhanced mel-spectrogram. We leverage a high-fidelity vocoder such as Vocos~\cite{siuzdak2023vocos} or BigVGAN~\cite{lee2022bigvgan}, which has demonstrated superior speech synthesis quality. The waveform reconstruction is given by:
\begin{equation}
    % \hat{x} = \text{Vocoder}(M_x),
    \hat{x} = \mathcal{V}(M_x),
\end{equation}
where $M_x$ is the enhanced mel-spectrogram predicted by the Flow Matching model.

By decoupling the enhancement process from waveform synthesis, FlowSE benefits from efficient training, as it operates directly in the mel-spectrogram domain while leveraging powerful pre-trained vocoders for waveform generation.

\vspace{-5pt}
\subsection{Training Objective}
\label{sec:training_objective}

FlowSE is trained using a flow-matching loss to learn the optimal velocity field for speech enhancement. Given a noisy-clean mel-spectrogram pair $(M_y, M_x)$ sampled from the data distribution $p(M_y, M_x)$, the objective is to model the continuous transformation from the noisy input to the clean target.

% \subsubsection{Flow Matching Loss}
% Following the Flow Matching framework~\cite{lipman2023flow, albergo2023stochastic}, we model the trajectory of the clean mel-spectrogram as a continuous flow from $M_y$ to $M_x$. The time-dependent state $M_t$ is defined as:
% \begin{equation}
%     M_t = \phi_t(M_y, M_x),
% \end{equation}
% where $\phi_t$ represents a deterministic interpolation function.

% The model learns the velocity field $v_\theta(M_t, t)$ that aligns with the optimal transport flow. The Flow Matching loss is formulated as:
% \begin{equation}
%     \mathcal{L}_{\text{FM}} = \mathbb{E}_{\substack{(M_y, M_x) \sim p(M_y, M_x) \\ t \sim \mathcal{U}(0,1)}} 
%     \left[ \lambda(t) \lVert v_\theta(M_t, t) - \tilde{v}(M_t, t) \rVert^2 \right],
% \end{equation}
% where $\tilde{v}(M_t, t)$ is the ground-truth velocity computed from the data trajectory, and $\lambda(t)$ is a weighting function that emphasizes different time steps.

% \subsubsection{Spectral Loss}
% To ensure that the enhanced speech maintains high perceptual quality, we introduce spectral loss that guides the model toward realistic speech reconstruction.

We incorporate an $\ell_1$ loss on the mel-spectrograms to directly minimize reconstruction error:
\begin{equation}
    \mathcal{L}_{\text{mel}} = \mathbb{E}_{(M_y, M_x) \sim p(M_y, M_x)} \left[ \lVert M_x - \hat{M}_x \rVert_1 \right],
\end{equation}
where $\hat{M}_x$ is the predicted clean mel-spectrogram.

% \subsubsection{Final Objective}

% The overall training objective combines the Flow Matching loss, spectral loss, and perceptual loss:
% \begin{equation}
%     \mathcal{L} = \mathcal{L}_{\text{FM}} + \alpha \mathcal{L}_{\text{mel}},
% \end{equation}
% where $\alpha$ is a hyperparameter set to 0.5.

% By jointly optimizing these losses, FlowSE learns to generate high-quality speech with accurate spectral content and perceptual fidelity.

\section{Experiments}
\label{sec:experiments}
\subsection{Datasets \& Evaluation Metrics}
\textbf{Training sets} 
To comprehensively evaluate the effectiveness of FlowSE, we construct a large-scale training dataset by combining multiple publicly available speech and noise datasets. We utilize WeNetSpeech~\cite{Yao2021WeNetPO}, GigaSpeech~\cite{chen2021gigaspeech}, VoiceBank(VCTK)~\cite{voicebank}, and the DNS Challenge - Interspeech 2021 datasets~\cite{reddy2020interspeech} to provide speech recordings. We incorporate environmental and artificial noise from the WHAM!~\cite{wichern2019wham}, DEMAND~\cite{thiemann2013diverse} and the DNS Challenge - Interspeech 2021 datasets~\cite{reddy2020interspeech}. We use room impulse responses from the OpenSLR26 and OpenSLR28 datasets~\cite{ko2017study}. Non-English and non-Chinese utterances are filtered out. Transcriptions are obtained using OpenAI's Whisper model~\cite{radford2022robustspeechrecognitionlargescale}\footnote{\url{https://huggingface.co/openai/whisper-large-v3-turbo}}.

\textbf{Test sets} 
For evaluation, we employ the publicly available DNS Challenge - Interspeech 2021 test set to compare FlowSE against state-of-the-art models. Additionally, we create a simulated test set by mixing clean speech from VCTK with unseen noise from WHAM! and DEMAND at various signal-to-noise ratios (SNRs) ranging from -5 dB to 10 dB.

\textbf{Evaluation Metrics}
It’s a known fact that standard metrics such as PESQ, SI-SNR
cannot accurately assess generative models due to a lack of waveform alignment. Therefore, we adopt DNSMOS, a neural network-based MOS estimator that closely correlates with human quality ratings. We use WeSpeaker~\cite{wang2023wespeaker} to compute speaker cosine similarity, assessing speaker identity preservation. Additionally, we employ OpenAI’s Whisper to transcribe enhanced speech and measure the word error rate (WER), which reflects intelligibility improvements. If necessary, generated speech is resampled to match the evaluation requirements.

\begin{table*}[t]
    \centering
    \caption{Comparison of different systems on the DNS Challenge test set.}
    \label{tab:results}
    \footnotesize  % 使用较小的字体以适应页面
    \begin{tabular}{c c ccccccccccc}
        \toprule
        \multirow{3}{*}{System} & \multirow{3}{*}{Model Type} & \multicolumn{4}{c}{With Reverb} & \multicolumn{4}{c}{Without Reverb} & \multicolumn{3}{c}{Real Recording} \\
        \cmidrule(lr){3-6} \cmidrule(lr){7-10} \cmidrule(lr){11-13}
         & & \multicolumn{3}{c}{DNSMOS $\uparrow$} & \multirow{2}{*}{Spk Sim $\uparrow$} & \multicolumn{3}{c}{DNSMOS $\uparrow$} & \multirow{2}{*}{Spk Sim $\uparrow$} & \multicolumn{3}{c}{DNSMOS $\uparrow$} \\
        \cmidrule(lr){3-5} \cmidrule(lr){7-9} \cmidrule(lr){11-13}
         & & SIG & BAK & OVRL &  & SIG & BAK & OVRL &  & SIG & BAK & OVRL \\
        \midrule
        Noisy                & --        & 1.760 & 1.497 & 1.392 & \textbf{0.941} & 3.392 & 2.618 & 2.483 & \textbf{0.969} & 3.053 & 2.510 & 2.255 \\
        \midrule
        TF-GridNet~\cite{wang2023tf} & Regression & 3.101 & 2.900 & 2.805 & 0.815 & 3.550 & 3.012 & 3.315 & 0.840 & 3.311 & 3.106 & 3.140 \\
        \midrule
        CDiffuSE~\cite{lu2022conditional}             & Diffusion & 2.541 & 2.300 & 2.190 & 0.741 & 3.294 & 3.641 & 3.047 & 0.765 & 3.201 & 3.104 & 2.781 \\
        SGMSE~\cite{welker2022speech}                & Diffusion & 2.730 & 2.741 & 2.430 & 0.764 & 3.501 & 3.710 & 3.137 & 0.782 & 3.297 & 2.894 & 2.793 \\
        StoRM~\cite{lemercier2023storm}                & Diffusion & 2.947 & 3.141 & 2.516 & 0.790 & 3.514 & 3.941 & 3.205 & 0.798 & 3.410 & 3.379 & 2.940 \\
        \midrule
        SELM~\cite{wang2024selm}                 & LM        & 3.160 & 3.577 & 2.695 & 0.793 & 3.508 & 4.096 & 3.258 & 0.810 & 3.591 & 3.435 & 3.124 \\
        MaskSR~\cite{li2024masksr}               & LM        & 3.531 & 4.065 & 3.253 & 0.827 & 3.586 & 4.116 & 3.339 & 0.929 & 3.430 & 4.025 & 3.136 \\
        \midrule
        FlowSE-w/ text     & Flow Matching        & \textbf{3.614} & \textbf{4.110} & \textbf{3.340} & 0.809 & \textbf{3.690} & 4.200 & \textbf{3.451} & 0.940 & \textbf{3.643} & \textbf{4.100} & \textbf{3.271} \\
        FlowSE-w/o text  & Flow Matching        & 3.601 & 4.102 & 3.331 & 0.801 & 3.685 & \textbf{4.201} & 3.445 & 0.934 & 3.635 & 4.080 & 3.263 \\
        \bottomrule
    \end{tabular}
\end{table*}

\vspace{-5pt}
\subsection{Training Setup}
We use 100-dimensional log mel-filterbank features extracted from 24 kHz audio with a hop length of 256 as input representation. We adopt a latent Diffusion Transformer (DiT)~\cite{peebles2023scalable} as our velocity field estimator. The model consists of $\text{N}$ = 22 layers, 16 attention heads, a hidden dimension of 1024, and a feed-forward network (FFN) dimension of 2048. We utilize ConvNeXt V2~\cite{woo2023convnext} as the text encoder, with an embedding dimension of 512 and an FFN dimension of 1024. It is trained to process character sequences extracted from transcriptions. A pre-trained Vocos vocoder~\cite{siuzdak2023vocos} is used to synthesize waveforms. We use the AdamW optimizer with a peak learning rate of $7.5 \times 10^{-5}$, which is linearly warmed up over 20K steps and then linearly decayed. A dropout rate of 0.1 is applied to attention layers and FFN layers. The gradient norm is clipped to a maximum value of 1 to stabilize training. The model is trained on 8 NVIDIA 4090D GPUs for 1.2 million updates.

\subsection{Baselines}
We compare FlowSE with state-of-the-art speech enhancement methods. TF-GridNet~\cite{wang2023tf} is a regression-based approach that integrates full- and sub-band modeling in the time-frequency domain. CDiffuSE~\cite{lu2022conditional} is a diffusion-based method operating in the time domain, while SGMSE~\cite{welker2022speech} adopts a score-based generative framework with a deep complex U-Net. StoRM~\cite{lemercier2023storm} also follows a diffusion-based strategy by incorporating a stochastic regeneration scheme. In contrast, SELM~\cite{wang2024selm} leverages a self-supervised speech representation model combined with a language model, and MaskSR~\cite{li2024masksr} utilizes a masked language model for speech enhancement.

We evaluate two variants of FlowSE: one that leverages transcription information (FlowSE-w/ text) and another that operates without text conditioning (FlowSE-w/o text). Official pre-trained models are used when available; otherwise, models are trained following their original settings. All systems employ the same Vocos vocoder to ensure a fair comparison.

% \vspace{-5pt}
\subsection{Experimental Results}

Table~\ref{tab:results} summarizes the performance of various systems on the DNS Challenge test set under three conditions: \emph{With Reverb}, \emph{Without Reverb}, and \emph{Real Recordings}. DNSMOS is used as a reference-free measure of perceptual quality, while speaker similarity (Spk Sim) evaluates how well the enhanced speech preserves the original speaker's characteristics. In addition, Table~\ref{tab:rtf_comparison} reports the real-time factor (RTF) and word error rate (WER) on a simulated test set, providing insights into computational efficiency and intelligibility.

\begin{table}[t]
\centering
\caption{\small Comparison of different systems on the simulated test set. Real-Time Factor (RTF) is measured on a single NVIDIA 4090D.}
\label{tab:rtf_comparison}
\footnotesize
\setlength{\tabcolsep}{4pt} % 调整列间距
\begin{tabular}{lccc}
\toprule
System          & Model Type    & WER $\downarrow$  & RTF $\downarrow$ \\ \midrule
Noisy                         & -                        & 28.10          & -    \\ \midrule
CDiffuSE~\cite{lu2022conditional}  & Diffusion              & 15.31          & 3.30 \\
SGMSE~\cite{welker2022speech}       & Diffusion              & 13.97          & 3.27 \\
StoRM~\cite{lemercier2023storm}     & Diffusion              & 14.00          & 3.49 \\ \midrule
FlowSE-w/ text                     & Flow Matching          & \textbf{8.79} & \textbf{0.31} \\
FlowSE-w/o text                     & Flow Matching          & 8.81 & \textbf{0.31} \\ \bottomrule
\end{tabular}
\end{table}

The baseline models, including the regression-based TF-GridNet, diffusion-based methods (CDiffuSE, SGMSE, StoRM) and LM-based approaches (SELM, MaskSR), show moderate improvements over the noisy baseline but still exhibit trade-offs between quality and speaker preservation. In particular, diffusion-based methods, despite competitive DNSMOS scores, suffer from lower speaker similarity.

In contrast, both variants of our proposed FlowSE deliver significant and consistent improvements across all test conditions. FlowSE with text conditioning (FlowSE-w/ text) achieves the highest DNSMOS scores, reaching up to 3.643 on \emph{Real Recordings}, and the best speaker similarity of 0.940. The text-free variant (FlowSE-w/o text) also performs competitively, demonstrating that our approach robustly enhances speech quality even in the absence of auxiliary transcription information. These results highlight FlowSE's ability to generate high-fidelity speech that not only sounds natural but also retains the speaker’s unique characteristics.

Furthermore, Table~\ref{tab:rtf_comparison} reveals a remarkable advantage of FlowSE in terms of computational efficiency. Diffusion-based methods exhibit high real-time factors (RTF above 3.0), reflecting the heavy computational load of their iterative denoising processes. In contrast, FlowSE achieves an RTF of 0.31—more than 10 times faster—while also significantly reducing the word error rate (WER) to approximately 8.8\%. The low WER indicates that FlowSE produces enhanced speech that is clearer and more intelligible, which is critical for downstream applications such as automatic speech recognition.

% Overall, our experimental results demonstrate that FlowSE sets a new state-of-the-art in speech enhancement. It not only achieves superior perceptual quality (as measured by DNSMOS) and preserves speaker identity (as shown by higher Spk Sim scores) but also ensures excellent semantic clarity (reflected by low WER) with a dramatic reduction in inference latency. These comprehensive advantages make FlowSE an ideal solution for real-time speech enhancement applications.

% \vspace{-5pt}
\section{Conclusion}
\label{sec:conclusion}
In this work, we introduce FlowSE, a novel speech enhancement model based on flow matching, which addresses the limitations of diffusion models' high complexity and slow inference, as well as the quantization loss in language-model-based approaches. FlowSE efficiently reconstructs high-quality speech while preserving speaker characteristics, achieving state-of-the-art performance. Our results highlight the potential of flow matching as a powerful alternative to existing generative methods for speech enhancement. We hope this work inspires further advancements in generative approaches for speech enhancement and restoration.

\clearpage

\bibliographystyle{IEEEtran}
\bibliography{mybib}

\end{document}